\documentclass[aps,prb,twocolumn,superscriptaddress,showpacs,floatfix]{revtex4-1}
\usepackage{hyperref}
\usepackage{amsmath}
\usepackage{graphicx}

\begin{document}

\title{New implementation of hybridization expansion quantum impurity solver based on Newton-Leja interpolation polynomial}
\author{Li Huang}
\affiliation{ Beijing National Laboratory for Condensed Matter Physics, 
              and Institute of Physics, 
              Chinese Academy of Sciences, 
              Beijing 100190, 
              China }
\affiliation{ Science and Technology on Surface Physics and Chemistry Laboratory, 
              P.O. Box 718-35, 
              Mianyang 621907, 
              Sichuan, 
              China }

\author{Xi Dai}
\affiliation{ Beijing National Laboratory for Condensed Matter Physics, 
              and Institute of Physics, 
              Chinese Academy of Sciences, 
              Beijing 100190, 
              China }

\date{\today}

\begin{abstract}
We introduce a new implementation of hybridization expansion continuous time quantum impurity 
solver which is relevant to dynamical mean-field theory. It employs Newton interpolation at a 
sequence of real Leja points to compute the time evolution of the local Hamiltonian efficiently. 
Since the new algorithm avoids not only computationally expansive matrix-matrix multiplications 
in conventional implementations but also huge memory consumptions required by Lanczos/Arnoldi 
iterations in recently developed Krylov subspace approach, it becomes advantageous over the
previous algorithms for quantum impurity models with five or more bands. In order to illustrate 
the great superiority and usefulness of our algorithm, we present realistic dynamical mean-field 
results for the electronic structures of representative correlated metal SrVO$_{3}$.
\end{abstract}

\pacs{02.70.Ss, 71.10.Fd, 71.30.+h, 71.10.Hf}

\maketitle

\section{introduction}
\label{sec:intro}

The rapid development of efficient numerical and analytical methods for solving quantum impurity 
models has been driven in recent years by the great success of dynamical mean-field theory 
(DMFT)\cite{antoine:13,kotliar:865} 
and its non-local extensions.\cite{maier:1027} In the framework of DMFT, the momentum dependence 
of self-energy is neglected, then the solution of general lattice model may be obtained from the 
solution of an appropriately defined quantum impurity model plus a self-consistency condition. 
Both the non-local extensions of DMFT\cite{maier:1027} and realistic DMFT (i.e, local density 
approximation combined with dynamical mean-field theory, LDA+DMFT) calculations,\cite{kotliar:865} 
involve multi-site or multi-orbital quantum impurity models, whose solutions are computationally 
expensive and in practice the bottleneck of the whole calculations. Therefore it is of crucial 
importance to develop fast, reliable, and accurate quantum impurity solvers.

The multi-site or multi-orbital nature of the most relevant quantum impurity models favors quantum
Monte Carlo methods. In the past two decades, perhaps the most commonly used impurity solver is 
the well-known Hirsch-Fye quantum Monte Carlo (HFQMC) algorithm.\cite{antoine:13,kotliar:865,maier:1027,hirsch:2521}
This solver is numerically exact, but computationally expensive. Furthermore it suffers inevitable 
systematic error which is introduced by time discretization procedure, and thus is not suitable 
for solving the quantum impurity models under low temperature. 

Very recently, important progresses have been achieved with the development of various continuous 
time quantum Monte Carlo (CTQMC) impurity solvers, which are based on stochastic sampling of 
diagrammatic expansion of the partition function.\cite{gull:349,gull:20111078} According to the 
differences in perturbation expansion terms, these CTQMC quantum impurity solvers can be classified 
into two types: weak coupling (also named as interaction expansion) and strong coupling (also 
named as hybridization expansion) implementations. The weak coupling CTQMC impurity solver 
was first proposed by Rubtsov \emph{et al.},\cite{rubtsov:035122,savkin:026402} who expanded the 
partition function in the interaction terms. This is the method of choice for cluster calculations 
of relatively simple models at small interactions, because the computational effort scales as the 
cube of the system size. As an useful complement, Werner and Millis proposed\cite{werner:155107,werner:076405} 
another powerful and flexible CTQMC impurity solver, which is based on a diagrammatic expansion in 
the impurity-bath hybridization and the local interactions are treated exactly. Since this algorithm 
perturbs around an exactly solved atomic limit, it is particularly efficient at moderate and strong 
interactions. Furthermore, due to its ability to provide the information about atomic states,\cite{haule:155113}
hybridization expansion quantum impurity solver is the desirable tool for LDA+DMFT calculations 
for strongly correlated materials. However, since the Hilbert space of the local problems grows 
exponentially with the number of sites or orbitals, the computational effort scales exponentially, 
rather than cubically with system size. Hence the applications of hybridization expansion quantum 
impurity solver in LDA+DMFT calculations are severely constrained, especially for those multi-orbital 
impurity models with rotationally invariant interaction terms. 

With this obstacle in mind, in this paper we present a new efficient implementation for hybridization 
expansion quantum impurity solver, which enables reliable and fast simulations for multi-orbital models 
with up to seven orbitals on modern computer clusters or GPU-enable workstations. The rest of this 
paper is organized as follows: In Sec.\ref{sec:hyb} a brief introduction to the conventional 
implementation\cite{werner:155107,haule:155113} and newly developed Krylov subspace 
approach\cite{lauchli:235117} for the hybridization expansion quantum impurity solver in general matrix 
formalism is provided. In Sec.\ref{sec:leja} the new algorithm based on Newton interpolation at real Leja 
points (for simplicity, in the following of this paper we just name it as Newton-Leja interpolation 
or Newton-Leja algorithm) is presented in details. Then in Sec.\ref{sec:bench} we discuss the truncation 
approximation, accuracy, and performance issues of the new algorithm and compare it with competitive 
Krylov subspace approach. In Sec.\ref{sec:app} the LDA+DMFT calculated results for typical strongly 
correlated metal SrVO$_{3}$ by using the Newton-Leja algorithm as an impurity solver are illustrated. 
Section \ref{sec:con} serves as a conclusion and outlook. In Appendix, concise introductions for the 
so-called Newton interpolation and Leja points are available as well.

\section{hybridization expansion quantum impurity solver}
\label{sec:hyb}

The general quantum impurity model to be solved in the framework of single site DMFT can be given by 
the following Hamiltonian:\cite{antoine:13,kotliar:865}
\begin{equation}
\label{eq:aim}
H_{qim}=-\sum_{a,\sigma}(\mu - \Delta_a) n_{a,\sigma} + H_{hyb} + H_{bath} + H_{int}.
\end{equation}
Here $a$ labels orbital, $\sigma$ labels spin, $\mu$ is the chemical potential, $\Delta_a$ is the 
energy level shift for orbital $a$ from a crystal field splitting, and $n_{a,\sigma}$ is the 
occupation operator. $H_{hyb}$, $H_{bath}$, and $H_{int}$ denote impurity-bath hybridization, bath environment, 
and interaction terms, respectively.

The basic idea of CTQMC impurity solvers is very simple.\cite{gull:349,gull:20111078} One begins 
from a general Hamiltonian $H= H_a + H_b$, which is split into two parts labeled by $a$ and $b$, writes the 
partition function $Z = \text{Tr} e^{-\beta H}$ in the interaction representation with respect to $H_a$, and 
expands in powers of $H_b$, thus
\begin{equation}
\label{eq:pert}
Z=\sum_k \frac{(-1)^{k}}{k!} (\prod^{k}_{i=1} \int^{\beta}_{0} d\tau_{i})
\text{Tr} [ e^{-\beta H_{a}} \mathcal{T} \prod^{k}_{i=1} H_{b} (\tau_{i})].
\end{equation}
Here $\mathcal{T}$ is the time-ordering operator.
The trace $\text{Tr}[...]$ evaluates to a number and diagrammatic Monte Carlo method enables a 
sampling over all orders $k$, all topologies of paths and diagrams, and all times $\tau_1$, 
$\cdots$, $\tau_k$ in the same calculation. Because this method is formulated in continuous time 
from the beginning, time discretization errors which are severe in Hirsch-Fye algorithm\cite{hirsch:2521} 
do not have to be controlled any more.

This continuous time method does not rely on an auxiliary field decomposition and a particular 
partitioning of the Hamiltonian into ``interacting" and ``non-interacting" parts. In principles, 
the only requirement is that one may decompose the Hamiltonian in such a way that the time 
evolution associated with $H_a$ and the contractions of operators $H_b$ may easily be evaluated. 
Thus there are several variations of CTQMC impurity solvers.\cite{gull:349,gull:20111078} We note 
that all of the continuous time diagrammatic expansion algorithms are based on the same general 
idea, there are only significant differences in the specifics of how the expansions are arranged, 
the measurements are done, and the errors are controlled. In this paper, we focus on the 
hybridization expansion algorithm merely.

In the hybridization expansion algorithm,\cite{werner:076405,werner:155107,haule:155113} based 
on Eq.(\ref{eq:aim}) and Eq.(\ref{eq:pert}), $H_b$ is taken to be the impurity-bath hybridization 
term $H_{hyb}$ and $H_a = H_{bath} + H_{loc}$, where
\begin{equation}
H_{loc} = H_{int} -\sum_{a,\sigma}(\mu - \Delta_a) n_{a,\sigma}.
\end{equation}
Since 
\begin{equation}
H_{hyb} = \sum_{pj}(V^{j}_{p} c^{\dagger}_{p} d_{j} + 
V^{j*}_{p}d^{\dagger}_{j}c_{p}) = \tilde{H}_{hyb} + \tilde{H}^{\dagger}_{hyb},
\end{equation}
contains two terms which create and annihilate electrons on the impurity, respectively,
only even powers of the expansion and contributions with equal numbers of $\tilde{H}_{hyb}$
and $\tilde{H}^{\dagger}_{hyb}$ can yield a nonzero trace. Inserting the $\tilde{H}_{hyb}$ 
and $\tilde{H}^{\dagger}_{hyb}$ operators explicitly into Eq.(\ref{eq:pert}) and
then separating the bath operators ($c^{\dagger}_p$ and $c_p$) and impurity operators 
($d^{\dagger}_j$ and $d_j$), we finally obtain
\begin{widetext}
\begin{equation}
\label{eq:zsep2}
\frac{Z}{Z_{bath}} = \sum_{k=0}^{\infty}
              \int^{\beta}_{0} d\tau_1 ... \int^{\beta}_{\tau_{k-1}} d\tau_k 
              \int^{\beta}_{0} d\tau'_1 ... \int^{\beta}_{\tau_{k'-1}} d\tau'_k 
              \sum_{\substack{b_1 ... b_k \\ b'_1 ... b'_k}}
              \text{Tr}_d \left[ e^{-\beta H_{loc}} 
              \mathcal{T} \prod^{k}_{i=1}
              d_{b_i}(\tau_i) d^{\dagger}_{b'_i}(\tau'_i) \right] 
              \text{det} \mathbf{\Delta}.
\end{equation}
\end{widetext}
Here we define the bath partition function
\begin{equation}
Z_{bath}= \text{Tr}e^{-\beta H_{bath}} = \prod_{\sigma} \prod_{p} (1 + e^{-\beta \epsilon_{p}}),
\end{equation}
and $\mathbf{\Delta}$ is a $k \times k$ matrix with elements $\mathbf{\Delta}_{lm} = 
\Delta_{j_l j_m}(\tau_l - \tau_m)$, where
\begin{equation}
\Delta_{lm}(\tau) = \sum_{\alpha} \frac{V^{l*}_{\alpha}V^{m}_{\alpha}}{1+e^{-\beta \epsilon_{\alpha}}}
\times
\begin{cases}
e^{\epsilon_{\alpha}(\tau - \beta)}, &\tau > 0 \\
- e^{\epsilon_{\alpha}\tau}, & \tau < 0.
\end{cases}
\end{equation}
Next the diagrammatic Monte Carlo technique can be used to sample Eq.(\ref{eq:zsep2}). The two 
basic actions required by ergodicity are the insertion and removal of a pair of creation 
and annihilation operators. Additional updates keeping the order $k$ constant are typical not 
required for ergodicity but may speed up equilibrium and improve the sampling efficiency.

In the Monte Carlo simulation, the most time consuming part is to calculate the following trace 
factor:\cite{gull:349,gull:20111078}
\begin{equation}
\label{eq:wloc}
w_{loc} = \sum_{\substack{b_1 ... b_k \\ b'_1 ... b'_k }}
              \text{Tr}_d \left[ e^{-\beta H_{loc}} 
              \mathcal{T} 
              \prod^{k}_{i=1} d_{b_i}(\tau_i) d^{\dagger}_{b'_i}(\tau'_i) \right].
\end{equation}
If $H_{loc}$ is diagonal in the occupation number basis defined by the $d^{\dagger}_{a}$ and 
$d_a$ operators, a separation of ``flavors" (spin, site, orbital, etc.) as in the segment 
formalism\cite{werner:076405} is possible, and the computational efficiency is fairly satisfactory. 
Conversely, if $H_{loc}$ is not diagonal in the occupation number basis, the calculation of 
$w_{loc}$ becomes more involved and challenging.

The conventional strategy proposed by Werner and Millis \emph{et al.}\cite{werner:155107,haule:155113} is 
to evaluate the trace factor (see Eq.(\ref{eq:wloc})) in the eigenstate basis of $H_{loc}$, because 
in this basis the 
time evolution operators $e^{-\tau H_{loc}}$ become diagonal and can be computed easily. On 
the other hand, in this representation the $d_{\alpha}$ and $d^{\dagger}_{\alpha}$ operators 
become complicated matrices. Hence this algorithm involves many multiplications of matrices
whose size is equal to the dimension of the Hilbert space of $H_{loc}$. In order to facilitate the task 
of multiplying these operator matrices it is crucial to arrange the eigenstates according to
some carefully chosen good quantum numbers.\cite{haule:155113} Then the evaluation of the trace 
is reduced to block 
matrix multiplications. With this trick, the present state of the art is that five spin 
degenerate bands can be treated exactly. However, since the matrix blocks are dense and the largest 
blocks grow exponentially with system size, the simulation of bigger models becomes extreme 
expensive and is only doable if the size of the blocks is severely truncated. Various truncation 
and approximation schemes provide limited access to larger problems, but as the number of 
orbitals is increased the difficulties rapidly become insurmountable.\cite{gull:349,gull:20111078}

Recently, L\"{a}uchli and Werner\cite{lauchli:235117} present an implementation of the hybridization
expansion impurity solver which employs sparse matrix exact diagonalization technique to compute
the time evolution of the local Hamiltonian $H_{loc}$ and then evaluate the weight of diagrammatic 
Monte Carlo configurations. They propose to adopt the occupation number basis, in which the creation 
and annihilation operators can easily be applied to any given states and in which the sparse nature 
of $H_{loc}$ matrix can be exploited during the imaginary time evolution by relying on mature Krylov 
subspace iteration method.\cite{park:5870,hoch:1911,moler:3} Their implementation is based on very 
efficient sparse matrix algorithm for the evaluation of matrix exponentials applied to a general vector, 
i.e., $\exp(-\tau H_{loc})|\nu\rangle$. At first the algorithm try to construct a Krylov subspace 
\begin{equation}
\mathcal{K}_{p}(|\nu\rangle) = \text{span}
\{|\nu\rangle,\ H_{loc}|\nu\rangle,\ H^{2}_{loc}|\nu\rangle,\ ...,\ H^{p}_{loc}|\nu\rangle\},
\end{equation}
by Lanczos or Arnoldi iterations, and then approximate the full matrix 
exponential of the Hamiltonian projected onto the Krylov subspace $\mathcal{K}_{p}(|\nu\rangle)$. 
Here $p$ means the dimension of the built Krylov subspace. It has been shown rigorously that these 
Krylov subspace iteration algorithms converge rapidly as a function of $p$.\cite{hoch:1911} Since 
this implementation involves only matrix-vector multiplications of the type $d^{\dagger} |\nu\rangle$, 
$d|\nu\rangle$, and $H_{loc}|\nu\rangle$ with sparse operators $d^{\dagger}$, $d$, and symmetric 
matrix $H_{loc}$, and is thus doable in principle even for systems for which the multiplication of dense 
matrix blocks becomes prohibitively expensive or for which the matrix blocks will not even fit 
into the memory anymore. Their algorithm avoids computationally expensive matrix-matrix multiplications 
and becomes advantageous over the conventional implementation for models with five or more bands.
La\"{u}chli \emph{et al.}\cite{lauchli:235117} have illustrated the power and usefulness of the Krylov 
subspace approach with dynamical mean-field results for a given five-band model which captures some 
aspects of the physics of the iron-based superconductors.

\section{Newton-Leja interpolation method}
\label{sec:leja}

The spirit of our new implementation for hybridization expansion quantum impurity solver is quite similar with 
previous Krylov subspace approach.\cite{lauchli:235117} We just adopt the occupation number basis 
and exploit the sparse nature of $d_{\alpha}^{\dagger}$ and $d_{\alpha}$ operators by applying 
them to any given states as well. But the kernel of our implementation is to evaluate the time evolution 
of sparse symmetric matrix $H_{loc}$, i.e. $\exp(-\tau H_{loc})$, by Newton interpolation at a sequence 
of real Leja points, 
instead of the Krylov subspace approach. Our algorithm inherits all the advantages of Krylov 
subspace approach, and is significantly superior to the latter on efficiency and memory consumption aspects. 
Consequently our implementation will be very promising to replace the Krylov subspace approach.

How to efficiently evaluate the matrix exponentials of local Hamiltonian $H_{loc}$ applied to any 
given vectors, i.e., $\exp(-\tau H_{lov})|\nu\rangle $, is the essential ingredient in hybridization 
expansion quantum impurity solver. We note that fast evaluation of the matrix exponential functions, just
like $\exp(\tau \mathcal{A}) v$ and $\varphi(\tau \mathcal{A}) v$, is the key 
building block of the so-called ``exponential integrators" in engineering mathematics and 
has received a strong impulse in recent years.\cite{hoch:209,hoch:1552,hoch:323} Here $\mathcal{A} 
\in \mathcal{R}^{n\times n}$, $v \in \mathcal{R}^{n}$, $\tau$ is arbitrary time step, and
\begin{equation}
\varphi(z) = \frac{\exp(z)-1}{z}.
\end{equation}
To this respect, 
most authors regard Krylov-like as the methods of choice.\cite{sidje:130} Nevertheless, an alternative
class of polynomial interpolation methods has been developed since the beginning,\cite{tal:25} which is based on 
direct interpolation or approximation of the matrix exponential functions on the eigenvalue spectrum (or the field 
of values) of the relevant matrix $\mathcal{A}$. Despite of a preprocessing stage needed to get a rough estimation of 
some marginal eigenvalues, the latter is competitive with Krylov-like methods in several instances, 
namely on large scale, sparse, and in general asymmetric matrices, usually arising from the spatial 
discretization of parabolic partial differential equations (PDEs).\cite{bcmv06}

Among others, the Newton interpolation based on real Leja points method (for a brief review of these two concepts, 
please refer to Appendix) has shown very attractive computational features.\cite{rei:332,bcmv06,mbcv09,cvb04,c07}
Given a general matrix $\mathcal{A}$ and a general vector $v$, Newton interpolation approximates 
the exponential propagator as 
\begin{equation}
\label{eq:appro}
\varphi(\tau \mathcal{A}) v \sim p_m (\tau \mathcal{A})v,
\end{equation}
with $m$ polynomial expansion order and $p_m (z)$ Newton interpolating polynomial of $\varphi(z)$ at a sequence of real Leja points 
$\{\xi_k\}$ in a compact subset of the complex plane containing the eigenvalue spectrum of matrix $\mathcal{A}$:\cite{rei:332,bcmv06,mbcv09,cvb04}
\begin{equation}
\label{eq:newton}
p_m (\tau \mathcal{A})v = \sum^{m}_{j=0} d_j \Omega_j v,
\end{equation}
and
\begin{equation}
\label{eq:omega}
\Omega_j = \prod^{j-1}_{k=0}(\tau \mathcal{A} - \xi_k \mathcal{I}).
\end{equation}
Here $\mathcal{I}$ is the unit matrix, and ${d_j}$ is the corresponding divided differences\cite{c07} for 
$\varphi(z)$. Observe that once $\varphi(\tau \mathcal{A})$ is computed, then 
\begin{equation}
\label{eq:trans}
\exp(\tau \mathcal{A})v = \tau \mathcal{A} \varphi(\tau \mathcal{A}) v + v.
\end{equation} 
Thus in practice it is numerically convenient to interpolate the function $\varphi(\tau \mathcal{A})$ at first.

By replacing general matrix $\mathcal{A}$ with $-H_{loc}$ and general vector $v$ with $|\nu \rangle$,
and using Eq.(\ref{eq:appro})-(\ref{eq:trans}), our algorithm rests on Newton interpolation of 
$\exp(-\tau H_{lov})|\nu\rangle $ at a sequence 
of Leja points on the real focal interval, say $[\alpha, \beta]$, of a suitable ellipse containing 
the eigenvalue spectrum of local Hamiltonian $-H_{loc}$. The use of real Leja points is suggested 
by the fact that on such a well defined ellipse they can give a stable interpolant, superlinearly 
convergent to entire functions due to the analogous scalar property,\cite{bag:124,cvb04} i.e.,
\begin{equation}
\label{eq:conv}
\lim_{m\rightarrow \infty} \sup || \varphi(\mathcal{A}) v - p_{m}(\mathcal{A})v||^{1/m}_{2} = 0.
\end{equation}
In real arithmetic, a key step is given by estimating at low cost the reference focal interval 
$[\alpha, \beta]$ for the eigenvalue spectrum of $-H_{loc}$ matrix. In present works we adopt 
the simplest estimation given directly by the famous Gershgorin's theorem.

Now let us describe the full workflow for the trace evaluation in some more details. In the 
initial stage of the hybridization expansion quantum impurity solver, the following steps are 
necessary: (1) Calculate the low-lying eigenstates of $H_{loc}$ by Lanczos iteration algorithm 
or exact diagonalization technique.\cite{antoine:13} Decide which eigenstates should be kept
in the trace calculation and the other high-lying eigenstates are discarded. (2) Determine 
the real focal interval $[\alpha,\beta]$ of a suitable ellipse containing the eigenvalue 
spectrum of matrix $-H_{loc}$ by using the Gershgorin's theorem. (3) Compute a sequence of real Leja 
points $\{\xi_k,\ k=0,...,m-1\}$ on the interval $[\alpha,\beta]$ by using the fast Leja points 
(FLPs) algorithm\cite{bag:124} 
and initialize the Newton-Leja algorithm. Depending on the simulation results gathered in 
present works the average degree for Newton interpolation $m \sim 15$, thus 64 real Leja 
points are enough to guarantee excellent convergence. Then, in the actual calculation of a trace factor, 
we proceed 
as follows: (4) Select an eigenstate as retained before, and propagate it to the first time evolution 
operator. Since the initial state is an eigenstate of $H_{loc}$, it is simply multiplied by 
an exponential factor for the first time interval. (5) Apply the creation or annihilation 
operator on the propagated state by using the efficient sparse matrix-vector multiplication 
technique. (6) Propagate the current state to next time evolution operator using the Newton-Leja 
algorithm as described above. (7) Go back to step 5 if more creation and annihilation operators 
are present. (8) Add the contribution of the propagated state to the trace. (9) Go back to 
step 4 until all retained eigenstates have been considered in the trace.

The Newton-Leja algorithm turns out to be quite simple and efficient, and its time complexity
is very similar with the Krylov subspace approach. According to Eq.(\ref{eq:newton}) and 
Eq.(\ref{eq:omega}), matrix-matrix multiplications are practically avoided, and the most 
important arithmetic is sparse matrix-vector multiplication. Furthermore, being based on 
vector recurrences in real arithmetic, its storage occupancy and computational cost are 
very small, and it results more efficient than Krylov subspace approach on large scale 
problems.\cite{bcmv06} In addition, this algorithm is very well structured for a parallel 
implementation, as it has been demonstrated in the references.\cite{mbcv09}
It is worth to mention that except for the traditional parallelism strategy for random walking and Markov chain 
in the Monte Carlo algorithm, in a fine-grained parallelism algorithm the (4) $\sim$
(8) steps can be easily parallelized over the retained eigenstates with multi-thread technique
in modern multi-core share memory computers. Further acceleration by using CUDA-GPU technology 
is another very promising research area. In the next section we should address the performance 
and reliability issues of this new algorithm. 

\section{benchmark}
\label{sec:bench}

In this section, we try to benchmark our new algorithm and compare the calculated results with 
other existing implementations for hybridization expansion quantum impurity solvers. 
Three aspects, including truncation error, reliability, and efficiency are mainly discussed.
Just like Krylov subspace approach, the truncation approximation can be adopted by our algorithm as well.
And the high accuracy and superior performance of our algorithm are proved with extensive test cases.

A spin degenerated three-band Hubbard model, with the following local Hamiltonian with rotationally 
invariant interactions\cite{gull:349}
\begin{equation}
\label{eq:loc}
\begin{split}
H_{loc} = &- \sum_{a,\sigma}(\mu - \Delta_a) n_{a,\sigma} + \sum_{a} Un_{a,\uparrow}n_{a,\downarrow}  \\
          &+ \sum_{a > b,\sigma}[U'n_{a,\sigma}n_{b,-\sigma} + (U'-J)n_{a,\sigma}n_{b,\sigma}] \\
          &- \sum_{a < b} J' (d^{\dagger}_{a,\downarrow}d^{\dagger}_{b,\uparrow}d_{b,\downarrow}d_{a,\uparrow} + h.c.) \\
          &- \sum_{a < b} J' (d^{\dagger}_{b,\uparrow}d^{\dagger}_{b,\downarrow}d_{a,\uparrow}d_{a,\downarrow} + h.c.),
\end{split}
\end{equation}
is used as a toy model to examine our implementation.
Here $U' = U -2 J$ and Hund's coupling parameter $J' = J = U/4$.  All the orbitals have equal bandwidth 4.0\ 
eV, and a semicircular density of states is chosen. The chemical potential $\mu$ is fixed to keep 
the system at half filling and the crystal field splitting $\Delta_{\alpha}$ is set to be zero. 
Unless it is specifically stated, this model is used throughout this section. We solve this toy 
model in the framework of single site DMFT using variant implementations of hybridization expansion 
quantum impurity solvers, including conventional matrix implementation,\cite{gull:349,gull:20111078} Krylov subspace
approach,\cite{lauchli:235117} and our Newton-Leja algorithm. 

\subsection{Truncation approximation}
\begin{figure}
\centering
\includegraphics[scale=0.60]{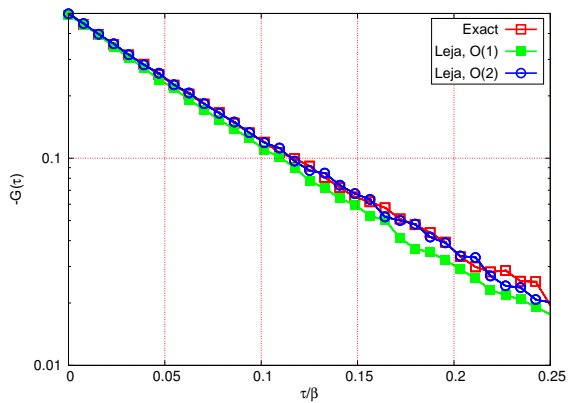}
\caption{(Color online) Single particle Green's function $G(\tau)$ of a three-band Hubbard 
model with $\beta = 10$ and $U$ = 2.0\ eV computed by the conventional matrix implementation 
(open squares, exact results) and the Newton-Leja algorithm (filled squares for $O(1)$ level 
truncation and open circles for $O(2)$ level truncation of the trace). \label{fig:trun}}
\end{figure}

In the calculation of trace factor, some high-lying eigenstates with negligible contributions 
can be abandoned in advance to improve the computational efficiency. At low temperature region 
this truncation approximation is practical, however, at high temperature region it must be used 
with great care. 

We solve the 
predefined three-band model (see Eq.(\ref{eq:loc})) at extreme high temperature ($\beta = 10$, $T \sim$ 1100\ K) and 
moderate interaction strength ($U$ = 2.0\ eV) to explore the influence of truncation approximation to 
the single particle Green's function. At first we can obtain the exact solutions by using the 
conventional matrix algorithm without
any truncations. Then we run the simulation again by using the Newton-Leja algorithm with $O(1)$ level 
(in which only the 4-fold degenerated ground states are retained) and $O(2)$ level (in which not only 
the 4-fold degenerated ground states but also the 28-fold degenerated first excited states are kept) 
truncations respectively. 

The calculated results are 
shown in Fig.\ref{fig:trun}. It is apparent that under $O(1)$ level truncation there are significantly 
systematic deviations between the approximate and exact results, while under $O(2)$ level truncation 
the deviations can be ignored safely. It has been suggested by La\"{u}chli \emph{et al.}\cite{lauchli:235117} 
that the truncation approximation to the ground state vectors is legitimate just for temperatures which 
are $\leq 1\%$ of the bandwidth. Therefore in this case the temperature is too high to apply the $O(1)$ 
level truncation, but the $O(2)$ level truncation is still acceptable. Finally, It should be pointed out 
that for this model the computational speed with $O(2)$ level truncation is at least twice faster than 
that of full calculation without any truncations.

\subsection{Accuracy of Newton-Leja algorithm}
\begin{figure}
\centering
\includegraphics[scale=0.60]{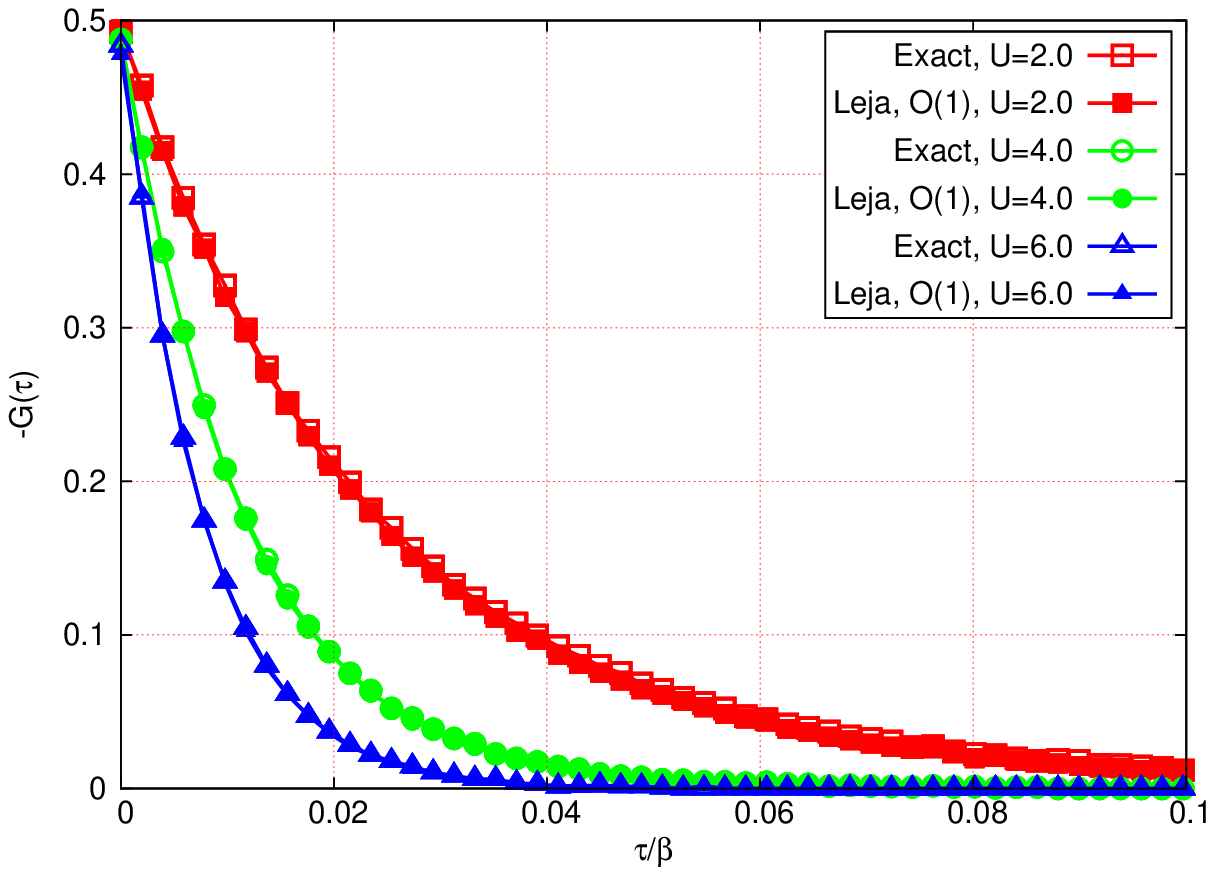}
\includegraphics[scale=0.60]{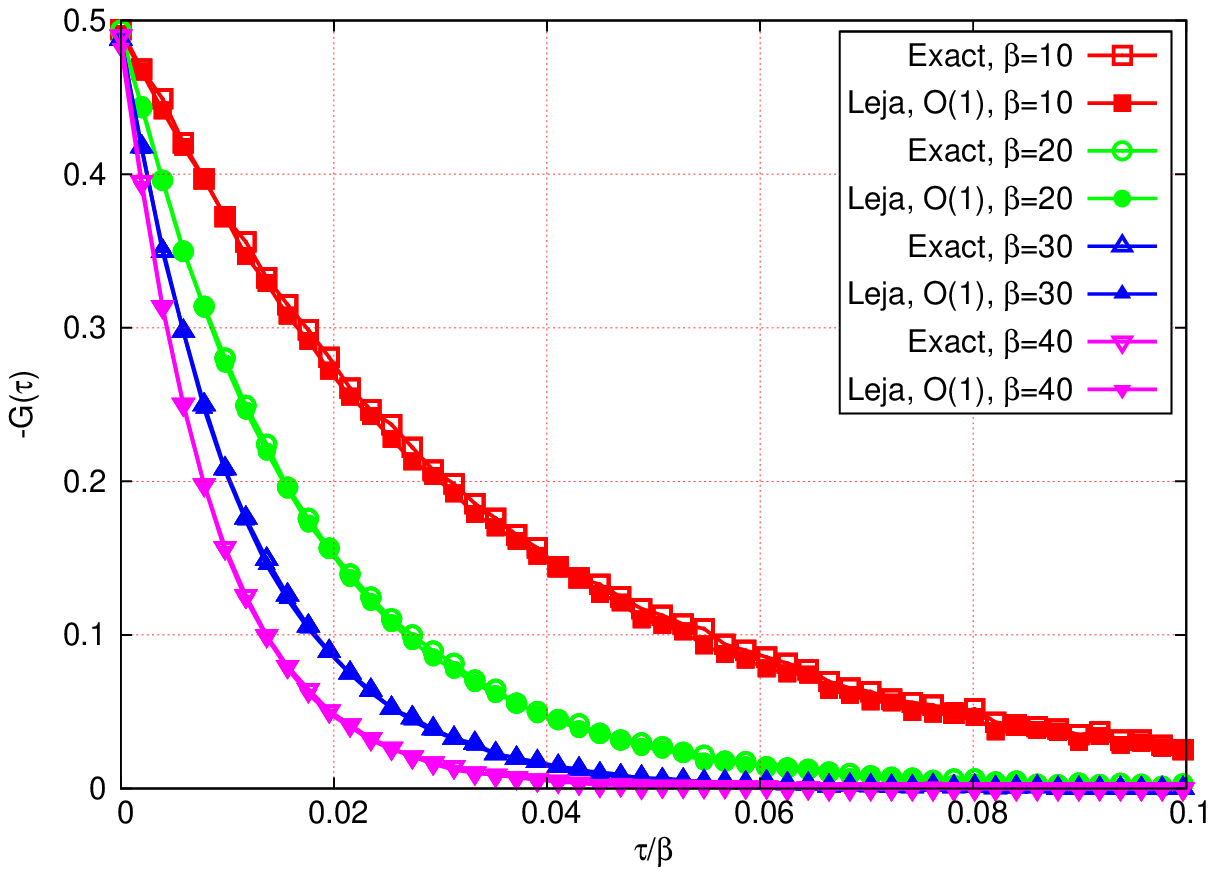}
\caption{(Color online) Comparison between the single particle Green's functions of a three-band model 
computed with the traditional matrix method (open symbols, without any truncations) and those computed 
with the Newton-Leja algorithm (full symbols, with $O(1)$ level truncation). Upper panel: the calculated 
single particle Green's functions at different interaction strengths $U$ and $\beta$ is fixed to 30. Lower 
panel: the calculated single particle Green's functions at $U$ = 4.0\ eV and different inverse temperatures $\beta$. 
\label{fig:g1_g2}}
\end{figure}

Next we try to demonstrate the accuracy of the new approach in a wider parameter range. The top panel 
of Fig.\ref{fig:g1_g2} shows the measured single particle Green's function $G(\tau)$ for $\beta = 30$ and different 
values of the Coulomb interaction strength $U$. Since the model temperature ($T \sim$ 390\ K) meets 
the truncation criterion,\cite{lauchli:235117}, both $O(1)$ and $O(2)$ level truncations are valid.
Of course, the $O(2)$ level truncation can give finer results at the cost of performance, so in these 
simulations we apply the $O(1)$ level truncation merely. In this figure the open 
symbols were computed with the conventional matrix method without any truncations and used 
to test the precision of the Newton-Leja algorithm. Not surprisingly, essentially perfect 
agreement between the two methods is found for all relevant interaction strengths.

The bottom panel of Fig.\ref{fig:g1_g2} illustrates the calculated results for $U$ = 4.0\ eV at different 
values of inverse temperature $\beta$. The $O(1)$ level truncation is adopted for the Newton-Leja algorithm. 
The exact solutions obtained by traditional matrix method are shown as a comparison. It is noticed that 
when the inverse temperature $\beta$ is 10, the deviations between traditional matrix implementation and Newton-Leja algorithm 
are distinct. The temperature is lower, the smaller the deviation. And for $\beta \geq 20 $, the deviations
can be considered negligible. In other words, their results become indistinguishable at temperatures which 
are $\leq 1\%$ of the bandwidth. Nevertheless, the Newton-Leja algorithm with carefully chosen approximations 
is demonstrated to be controllable, reliable and consistent with original implementations, and can be widely used in
standard DMFT calculations.

\subsection{Efficiency of Newton-Leja algorithm}

Efficiency is always a major concern for newly developed quantum impurity solver, so it is urgent for us to 
explore the performance of Newton-Leja algorithm. Based on considerable testing, we found that 
when the system size is not large enough, the Newton-Leja algorithm is less efficiency than general 
matrix formulation,\cite{gull:349,gull:20111078} even the Krylov subspace approach.\cite{lauchli:235117} 
Whereas, when the system size is large enough, e.g., five-band or bigger model, the situation completely opposite. 
The Krylov subspace approach outperforms the conventional matrix method and Newton-Leja algorithm is better than 
the Krylov subspace approach. In this subsection, we try to compare the efficiencies between Newton-Leja 
algorithm and Krylov subspace approach, and determine the system size for which the former is superior 
to the latter. 

\begin{figure}
\centering
\includegraphics[scale=0.60]{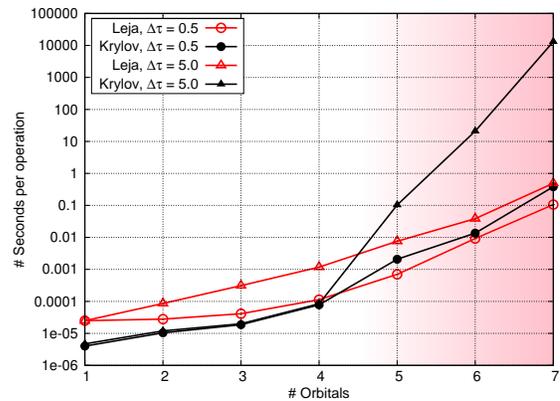}
\caption{(Color online) Efficiencies (time consumption per evaluation of matrix exponentials 
$\exp(-\Delta\tau H_{loc})|\nu\rangle$ is used as metric) of the Krylov subspace approach and 
Newton-Leja algorithm as a function of system size. The local Hamiltonian $H_{loc}$ with 
rotationally invariant interaction at half-filling is defined in Eq.(\ref{eq:loc}), where 
$U$ = 4.0\ eV and $J/U = 0.25$. $\Delta\tau$ = 0.5 and 5.0 for typically short and long time 
intervals, respectively. The initial vector $|\nu\rangle$ is generated randomly. Both the 
maximum number of real Leja points for Newton-Leja algorithm $m$ and maximum allowable dimension 
of Krylov subspace for Krylov approach $p$ are 64. The results obtained by these two algorithms 
are consistent with each other within the machine precision. In the pink zone, the efficiency 
of Newton-Leja algorithm is clearly superior to Krylov subspace approach.\label{fig:time}}
\end{figure}

Since the key building blocks for both Newton-Leja algorithm and Krylov subspace approach are the
calculations of time evolution operators, we just compare their efficiencies by the evaluation of 
$\exp(-\Delta\tau H_{loc}) |\nu\rangle$. Here $|\nu\rangle$ is a randomly generated vector and time 
step $\Delta\tau$ is set to be 0.5 and 5.0, corresponding to typically short and long time intervals, 
respectively. The multi-orbital local Hamiltonian $H_{loc}$ is constructed by using Eq.(\ref{eq:loc}), 
and the interaction strength $U$ is fixed to be 4.0\ eV and $J' =J = U/4$. Both the maximum allowable 
dimension for Krylov subspace $p$ and number for real Leja points $m$ are fixed to be 64, which are 
sufficient to obtain convergent and accurate results.

In order to eliminate the influence of fluctuating measurement data, we repeated every benchmark for 20
times and then evaluated the average time consumption per evaluation of matrix exponentials. The benchmark 
results for multi-orbital systems with $n=$1, 2, $\cdots$, 7 ($n$ labels the number of bands) 
are displayed in Fig.\ref{fig:time}. As can be seen from the figure, when the system size is 
small or moderate ($n \leq 4$) the Newton-Leja algorithm exhibits worse performance than Krylov 
subspace approach. However, when increasing the system size continually ($ 5 \leq n \leq 7 $, 
indicated in this figure by pink zone) the Newton-Leja algorithm is the winner. For instance, 
at $n=5$ and $\Delta\tau=5.0$ the Newton-Leja algorithm is almost ten times faster than the 
Krylov subspace approach. More impressively, at $n=7$ and $\Delta \tau=5.0$, the Krylov subspace approach 
is slower than Newton-Leja algorithm about even four orders of magnitude. We note that for short time 
interval the performances for both implementations are close, but as for $n \geq 5 $ the Newton-Leja 
algorithm still exhibits better performance. According to these benchmarks, it is tentatively 
suggested that the Newton-Leja algorithm is much more suitable for the systems with five or more 
orbitals than Krylov subspace approach.

\begin{figure}
\centering
\includegraphics[scale=0.60]{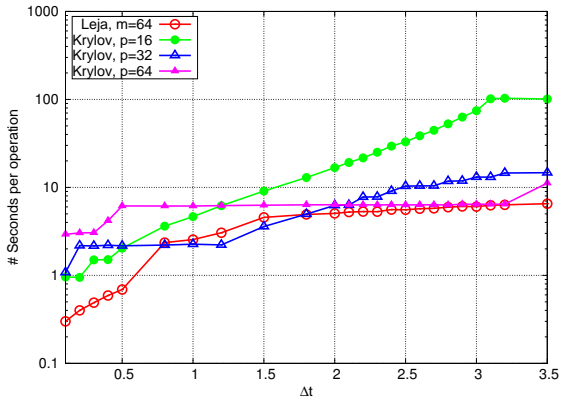}
\caption{(Color online) Efficiencies (time consumption per evaluation of matrix exponentials 
$\exp(-\Delta\tau H_{loc})|\nu\rangle$ is used as metric) of the Krylov subspace approach and 
Newton-Leja algorithm as a function of time interval for the five-band 
model with rotationally invariant interaction ($U=$ 4.0\ eV, $J/U=0.25$). The chemical 
potential $\mu$ is fixed to fulfill the half-filling condition. Randomly generated vector 
$|\nu\rangle$ is used as an initial state. The maximum number of real Leja points $m$ is fixed 
to be 64, while the maximum allowable size of Krylov subspace $p$ is varied from 16 to 64. The 
convergence criterion for all algorithms are the same. When $\Delta\tau > 3.5$, it is very 
difficult to obtain converged solutions for Krylov subspace approach by current settings.\label{fig:5t}}
\end{figure}

Next we concentrate our attentions to the five-band model, which plays an important role in the 
underlying physics of transition metal compounds, and make further benchmarks for our new 
implementation. The average time consumption per evaluation of matrix exponentials 
$\exp(-\Delta\tau H_{loc})|\nu\rangle$) is used again as a measurement of efficiency. The 
model parameters are consistent with previous settings except the time interval $\Delta\tau=0.1 
\sim 3.5$. The maximum number of real Leja points $m$ is fixed to be 64, while the maximum allowable 
dimension of Krylov subspace $p$ is varied from 16 to 64. The benchmark results are shown in 
Fig.\ref{fig:5t}. As is illustrated in this figure, at short ($\Delta\tau < 0.8$) and long 
($\Delta\tau > 1.8 $) time intervals, the Newton-Leja algorithm shows better efficiency. 
It is apparent that when $\Delta\tau > 2.0$, actually no decline in efficiency for Newton-Leja algorithm 
is seen, i.e., the efficiency has nothing to do with the length of time interval. That is 
because the average degree of Newton interpolation (i.e. number of Leja points used in Newton interpolation) almost 
remain unchanged. According to our experiences, $m = 15 \sim 20$ is suitable for most of the 
five, six, and seven-band models in general parameter ranges. As is mentioned above, provided 
limiting $m$ and a well defined ellipse containing the eigenvalue spectrum of matrix $-H_{loc}$, a 
stable interpolant, superlinearly convergent to matrix exponentials $\exp(-\Delta\tau H_{loc})|\nu\rangle$) 
is guaranteed by the Leja points method (see Eq.(\ref{eq:conv})).\cite{bag:124,cvb04}
So we can decrease the maximum number of real Leja points $m$ further to reduce the memory 
consumption and obtain higher efficiency. 
As for the Krylov subspace approach, at short time interval region the simulation with 
smaller Krylov subspace exhibits better performance, while at long time interval region the contrary 
is indeed true. It is very easy to be understood: at long time interval region, the Krylov subspace algorithm 
requires larger dimension of subspace to obtain fast convergence speed, and it is hardly 
to achieve convergence with small dimension of Krylov subspace only when more iterations are 
done. However, it is impossible to increase the dimension of Krylov subspace infinitely, the 
oversize of Krylov subspace will deteriorate the performance of Krylov approach quickly. 

\begin{table}
\caption{The average computational times per DMFT iteration for typical three-band and five-band Hubbard 
models solved by hybridization expansion quantum impurity solvers based on Newton-Leja algorithm and 
Krylov subspace approach respectively. In current simulations, each DMFT iteration took 40000000 Monte Carlo steps
per processor, and the results were averaged over 16 processors.\label{tab:time}}
\begin{tabular}{ccc}
\hline \hline
method & three-band model & five-band model\\
\hline
& \multicolumn{2}{c}{$U = 4.0$\ eV and $\beta = 20$} \\
Newton-Leja, $m = 64$  & 3.23h   & 29.07h  \\
Krylov, $p = 64$       & 0.46h   & 36.80h  \\
\hline
& \multicolumn{2}{c}{$U = 4.0$\ eV and $\beta = 40$} \\
Newton-Leja, $m = 64$  & 6.12h   & 67.32h  \\
Krylov, $p = 64$       & 0.78h   & 399.36h \\
\hline
\hline
\end{tabular}
\end{table}

Finally, two concrete cases are provided to demonstrate the superior performance of Newton-Leja
algorithm. Let's consider typical three-band and five-band Hubbard models respectively. The 
moderate Coulomb interaction strength $U = 4.0$ eV, and inverse temperature 
$\beta = 20\ \text{or}\ 40$ ($T \sim$ 580\ K or 290\ K) are 
chosen. These two models are solved separately by hybridization expansion quantum impurity solvers 
based on Newton-Leja algorithm and Krylov subspace approach with the same computational parameters, 
and the computational times are gathered and compared with each other. The benchmark results are 
summarized in table \ref{tab:time}. It is confirmed again that for the three-band model Krylov subspace 
approach is more efficient than Newton-Leja algorithm, yet for the five-band model Newton-Leja 
algorithm exhibits much better performance, which is consistent with previous benchmark results.
Thus by any measure, our new implementation is better than Krylov subspace approach at 
low temperature and large system size.

\section{application}
\label{sec:app}

In this section, in order to illustrate the usefulness of Newton-Leja algorithm, we used it as 
a quantum impurity solver in the non-self-consistent LDA+DMFT calculations for representative 
transition metal oxide SrVO$_{3}$. 
SrVO$_{3}$ is a well-known $t^{1}_{2g}e^{0}_{g}$ metal. It is a good test case for LDA+DMFT 
calculations because it is cubic and nonmagnetic and also the $t_{2g}$ bands are isolated 
from both $e_g$ and oxygen $2p$ bands in the LDA band structure. Numerous theoretical calculations
(including LDA+DMFT)\cite{pav:176403,nek:155112,amadon:205112,nek:155106} and experiments\cite{egu:076402,
yos:146404,sek:156402,yos:085119} have been done on this compound. This is thus an ideal system
to benchmark our implementation.

\begin{figure}
\centering
\includegraphics[scale=0.60]{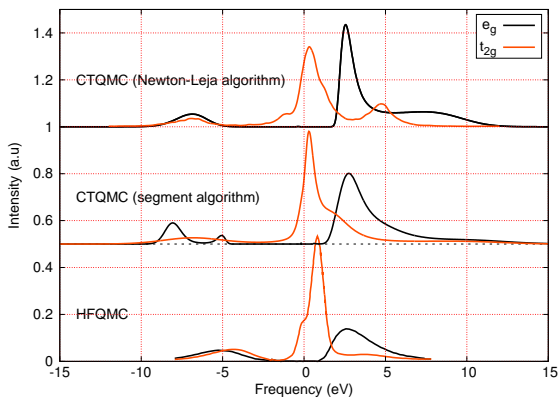}
\caption{(Color online) The single particle spectral functions for vanadium $3d$ states in
SrVO$_{3}$ obtained by LDA+DMFT calculations. As for the Newton-Leja algorithm, the rotationally 
invariant interaction terms are taken into considerations, while for segment algorithm\cite{werner:076405} 
only the Ising-like density-density interaction terms are treated. The spectral function is 
obtained from imaginary time Green's function $G(\tau)$ by using maximum entropy method,\cite{jarrell:133} 
and the calculated results are cross-checked by using recently developed stochastic 
analytical continuation method.\cite{beach} Previous LDA+DMFT results,\cite{amadon:205112} in 
which traditional HFQMC is used as an impurity solver, are shown in this figure as a 
comparison. \label{fig:spectrum}}
\end{figure}

The LDA+DMFT framework employed in present works has been described in the 
literatures.\cite{korotin:91,trimarchi:135227,amadon:205112} The ground state calculations 
have been carried out by using the projector augmented wave (PAW) method with the {\scriptsize 
ABINIT} package. The cutoff energy for plane wave expansion is 20 Ha, and the $k$-mesh for Brillouin zone
integration is $12 \times 12 \times 12$. The low-energy effective LDA Hamiltonian is obtained by applying a projection 
onto maximally localized Wannier function (MLWF) orbitals including all the vanadium $3d$ and 
oxygen $2p$ orbitals, which is described in details in reference [\onlinecite{amadon:205112}]. 
That would correspond to a $14 \times 14$ $p-d$ Hamiltonian which is a minimal model\cite{mos:033104} 
required for a correct description of the electronic structure of SrVO$_{3}$.
 
The LDA+DMFT calculations presented below have been done for the experimental lattice 
constants ($a_{0}=7.2605$ a.u). 
All the calculations were preformed in paramagnetic state at the temperature of 1160\ K ($\beta=10$).
The Coulomb interaction is taken into considerations merely among vanadium $3d$ orbitals. In the 
present work, we choose $U$ = 4.0\ eV and $J$ = 0.65\ eV, which are accordance to previous LDA+DMFT
calculations.\cite{amadon:205112} We adopt the around mean-field (AMF) scheme proposed in reference 
[\onlinecite{amadon:205112}] to deal with the double counting energy. The effective quantum impurity problem 
for the DMFT part was solved by two implementations of hybridization expansion CTQMC quantum impurity solver supplemented 
with recently developed orthogonal polynomial representation algorithm.\cite{boe:075145} The former
implementation is based on the segment representation\cite{werner:076405} and only the Ising-like
density-density interaction terms are treated. The latter implementation is based on the Newton-Leja 
algorithm, and the local Hamiltonian is with rotationally invariant interactions (see Eq.(\ref{eq:loc})). 
The maximum entropy method\cite{jarrell:133} was used to perform analytical continuation to obtain 
the impurity spectral function from imaginary time Green's function $G(\tau)$ of vanadium $3d$ states.
In the present simulations, each LDA+DMFT iteration took 40000000 Monte Carlo steps per process.
Since the segment representation hybridization expansion impurity solver\cite{werner:076405} is 
extreme efficient, it took less 5 hour to finish 30 LDA+DMFT iterations by using a 8-cores 
Xeon CPU. Though $O(2)$ level approximation is adopted during the simulation, the Newton-Leja 
algorithm is much more time consumption and took about $12 \sim 13$ hours to finish single LDA+DMFT 
iteration by using 64 cores in a Xeon cluster.

The calculated single particle spectral functions for vanadium $3d$ states in SrVO$_{3}$ are 
shown in Fig.\ref{fig:spectrum}. The calculated results within density-density interaction are 
consistent with previous LDA+DMFT simulations.\cite{amadon:205112,nek:155112,nek:155106} But
the results obtained by classical segment representation and Newton-Leja algorithm display 
remarkable differences. For examples, the upper and lower Hubbard bands in $t_{2g}$ sub-bands 
obtained by Newton-Leja algorithm with rotationally invariant interactions are more apparent. 
And the quasiparticle resonance peak around the Fermi level exhibits clearly shoulder structure, 
while that shoulder peak is absent in the segment picture (density-density interaction case). 
Since the vanadium $e_{g}$ sub-bands are isolated from $t_{2g}$ states and located above the 
Fermi level, so they show roughly similar peak structures for both two implementations of hybridization 
expansion quantum impurity solvers. Nevertheless, based on our calculated results, it is suggested that 
the spin-flip and pair-hopping terms in rotationally invariant interactions may play a key role 
in understanding the subtle electronic structure of SrVO$_{3}$ around the Fermi level, which has 
been ignored by previous theoretical calculations.

\section{conclusions}
\label{sec:con}
We have presented an alternate implementation of the hybridization expansion quantum impurity 
solver which makes use of the Newton-Leja interpolation to evaluate the weight of Monte Carlo 
configurations. The new implementation inherits all the advantages of previously developed Krylov 
subspace approach with less memory consumption and better convergence control. It shows tremendous 
growth in computational performance over Krylov subspace approach and conventional matrix implementation
at low temperature and large system size, and provides a controlled and efficient way which 
enables the LDA+DMFT study of transition metal compounds, or even actinide compounds with realistic 
interactions. To demonstrate the power of the new implementation, we used it as an impurity solver 
in the LDA+DMFT calculations for typical strongly correlated metal SrVO$_{3}$. The obtained impurity 
spectral function of vanadium $t_{2g}$ states shows apparent distinctions with previous calculated 
results. It is argued that the full Hund's exchange may has a big impact on the fine energy spectrum
near the Fermi level. The generalization of Newton-Leja algorithm to high performance CUDA-GPU 
architecture is underway.

\begin{acknowledgments}
We acknowledge financial support from the National Science Foundation ͑of China and that
from the 973 program of China under Contract No.2007CB925000 and No.2011CBA00108. All the 
LDA+DMFT calculations have been performed on the SHENTENG7000 at Supercomputing Center of 
Chinese Academy of Sciences (SCCAS).
\end{acknowledgments}

\appendix
\section{Leja points}
\begin{figure}
\centering
\includegraphics[scale=0.60]{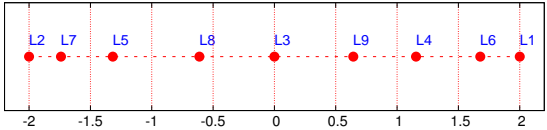}
\caption{(Color online) The first nine asymmetric real Leja points in [-2:2] interval generated by ``Fast 
Leja Points" algorithm.\cite{bag:124} The labels denote the sequences of Leja points.\label{fig:lp}}
\end{figure}
For the reader's convenience, here we briefly recall the definition of real Leja points. 
Sequences of Leja points $\{\xi_j\}^{\infty}_{j=0}$ for the compact set $\mathbf{K}$ are defined recursively as follows: 
if $\xi_0$ is an arbitrary fixed point in $\mathbf{K}$, then $\xi_j$ are generated recursively in 
such a way that
\begin{equation}
\prod^{j-1}_{k=0}|\xi_j - \xi_k| = \max_{\xi \in \mathbf{K}} \prod^{j-1}_{k=0} |\xi - \xi_k|,\ j = 1,\ 2,\ \cdots.
\end{equation}
An efficient algorithm for computing a sequence of real Leja points, the so-called ``Fast Leja Points 
(FLPs)", has been proposed in reference [\onlinecite{bag:124}]. In Fig.\ref{fig:lp} the first nine Leja points 
in [-2,2] interval are shown as a illustration.
The Leja sequences are attractive for interpolation at high-degree, in view of the stability of the 
corresponding algorithm in the Newton interpolation.\cite{rei:332}

\section{Newton interpolation}
Given a set of $m + 1$ data points
\begin{equation}
    (x_0, y_0), \ldots, (x_m, y_m)
\end{equation}
where no two $x_j$ are the same, the so-called Newton interpolation polynomial is defined as follows
\begin{equation}
\label{eq:a2}
p_m (x) = \sum^{m}_{j=0} d_j \Omega_j (x),
\end{equation}
with ${d_j}$ the divided difference\cite{c07} and the Newton basis polynomial $\Omega_j(x)$ defined as
\begin{equation}
\label{eq:a3}
\Omega_j(x) = \prod^{j-1}_{k=0} (x - x_k).
\end{equation}
It is apparent that Eq.(\ref{eq:newton}) and Eq.(\ref{eq:omega}) are the 
matrix forms of Eq.(\ref{eq:a2}) and Eq.(\ref{eq:a3}), respectively.
The Newton form is an attractive representation for interpolation polynomials
because it can be determined and evaluate rapidly, and, moreover, it is
easy to determine $p_{m+1}$ if $p_{m}$ is already known.
Because the Leja points are defined recursively, they are attractive to use with the Newton interpolation. 

\bibliography{leja}

\end{document}